\pdfoutput=1 % Added for astro-ph submission
\documentclass[]{spie}  %>>> use for US letter paper
\usepackage[]{graphicx}
\usepackage{deluxetable}
\usepackage{amsmath}
\usepackage{color}
\title{The Habitable-Zone Planet Finder: A Stabilized Fiber-Fed NIR Spectrograph for the Hobby-Eberly Telescope } 

\author{Suvrath Mahadevan\supit{a,b}, Lawrence Ramsey\supit{a,b}, Chad Bender\supit{a,b}, Ryan Terrien\supit{a,b}, Jason  T. Wright\supit{a,b}, Sam Halverson\supit{a,b}, Fred Hearty\supit{c},  Matt Nelson\supit{c}, Adam Burton\supit{c},  Stephen Redman\supit{d}, Steven Osterman\supit{e}, Scott Diddams\supit{f}, James Kasting \supit{a,b}, Michael Endl\supit{g}, Rohit Deshpande\supit{a,b}
\skiplinehalf
\supit{a}Department of Astronomy \& Astrophysics, The Pennsylvania State University, 525 Davey Laboratory, University Park, 16802, USA; \\
\supit{b}Center for Exoplanets \& Habitable Worlds, The Pennsylvania State University, University Park, PA 16802; \\
\supit{c}Department of Astronomy, University of Virginia, P.O. Box 400325, Charlottesville, VA 22904-4325, USA; \\
\supit{d}Atomic Physics Division, National Institute of Standards and Technology, Gaithersburg, MD 20899, USA; \\
\supit{e} Center for Astrophysics and Space Astronomy, University of Colorado, Boulder, CO; \\
\supit{f} National Institute of Standards and Technology, 325 Broadway, Boulder, CO; \\
\supit{g} McDonald Observatory, The University of Texas at Austin, Austin, Texas 78712, USA; \\
}

\authorinfo{E-mail: suvrath@psu.edu, Telephone: 1 814 865 0361}
\begin{document} 
  \maketitle 

%%%%%%%%%%%%%%%%%%%%%%%%%%%%%%%%%%%%%%%%%%%%%%%%%%%%%%%%%%%%% 
\begin{abstract}
We present the scientific motivation and conceptual design for the recently funded Habitable-zone Planet Finder (HPF), a stabilized fiber-fed near-infrared (NIR) spectrograph for the 10 meter class Hobby-Eberly Telescope (HET) that will be capable of discovering low mass planets around M dwarfs. The HPF will cover the NIR Y \& J bands to enable precise radial velocities to be obtained on mid M dwarfs, and enable the detection of low mass planets around these stars. The conceptual design is comprised of a cryostat cooled to 200K, a dual fiber-feed with a science and calibration fiber, a gold coated mosaic echelle grating, and a Teledyne Hawaii-2RG (H2RG) \footnote[1]{Certain commercial equipment, instruments, or materials are identified in this paper in order to specify the experimental procedure adequately. Such identification is not intended to imply recommendation or endorsement by the National Institute of Standards and Technology, nor is it intended to imply that the materials or equipment identified are necessarily the best available for the purpose.}  NIR detector with a 1.7$\mu$m cutoff. A uranium-neon hollow-cathode lamp is the baseline wavelength calibration source, and we are actively testing laser frequency combs to enable even higher radial velocity precision. We will present the overall instrument system design and integration with the HET, and discuss major system challenges, key choices, and ongoing research and development projects to mitigate risk. We also discuss the ongoing process of target selection for the HPF survey.
\end{abstract}

%>>>> Include a list of keywords after the abstract 


%%%%%%%%%%%%%%%%%%%%%%%%%%%%%%%%%%%%%%%%%%%%%%%%%%%%%%%%%%%%%
\section{INTRODUCTION}
\label{sec:intro}  % \label{} allows reference to this section

Considerable interest is now focused on finding and characterizing terrestrial-mass planets in habitable zones around their host stars. Such planets are extremely difficult to detect around F, G, and K stars, requiring either very high radial velocity (RV) precision ($\ll 1$ m/s) or space-based photometry. The Kepler mission  \cite{2010Sci...327..977B} aims to discover the signature of transiting earth-like planets, but confirmation with radial velocity remains very challenging. Efforts are beginning to focus on M dwarfs as their lower luminosity shifts the habitable zone (HZ \cite{1993Icar..101..108K}, Figure 1, a region around a star where liquid surface water may exist on a planet) much closer to the star. The lower stellar mass of the M dwarfs, as well as the short orbital periods of HZ planets, increases the Doppler wobble caused by a terrestrial-mass planet. RV studies have uncovered planetary systems around $\sim 20$ M dwarfs to date, including the low mass planetary system around GJ581 \cite{2009A&A...507..487M}, and KOI-961 \cite{2012ApJ...747..144M}. These observations suggest that, while hot Jupiters may be rare in M star systems \cite{2006ApJ...649..436E}, lower mass planets do exist around M stars and may be rather common. Theoretical work based on core-accretion models and simulations also predicts that short period Neptune mass planets should be common around M stars \cite{2005ApJ...626.1045I}. Climate simulations of planets in the HZ around M stars \cite{1997Icar..129..450J} show that tidal locking does not necessarily lead to atmospheric collapse The habitability of terrestrial planets around M stars has also been explored by many groups \cite{2007AsBio...7...85S}.  
As seen in Figure 1,  10 Earth-mass planets in the HZ are already detectable at more than $3\sigma$ with a velocity precision of 3m/s, and an instrument capable of 1-3m/s precision will have the sensitivity to discover terrestrial mass planets around the majority of mid-late M dwarfs.  Such precision is already achievable with high-resolution optical echelle spectrographs \cite{1996PASP..108..500B, 2012arXiv1206.5307B}. However, nearly all the stars in current optical RV surveys  are earlier in spectral type than $\sim$M4 since later spectral types are difficult targets even on large telescopes due to their intrinsic faintness in the optical: they emit most of their flux in the NIR between 0.9 and 1.8 $\mu$m (the Y, J and H bands, 0.98-1.1 $\mu$m, 1.1-1.4 $\mu$m and 1.45-1.8 $\mu$m). However, it is the low mass late-type M stars, which are the least luminous, where the velocity amplitude of a terrestrial planet in the habitable zone is highest, making them very desirable targets. Since the flux distribution from M stars peaks sharply in the NIR \cite{2006A&A...447..709P}, a stable high-resolution NIR spectrograph capable of delivering high RV precision can observe several hundred of the nearest M dwarfs to examine their planet population.  {\bf HPF is being designed to be such a stable  NIR spectrograph, with its primary science goal being the search for planets around mid-late M dwarfs.}

\begin{figure}
\begin{center}
\includegraphics[width=5.5in]{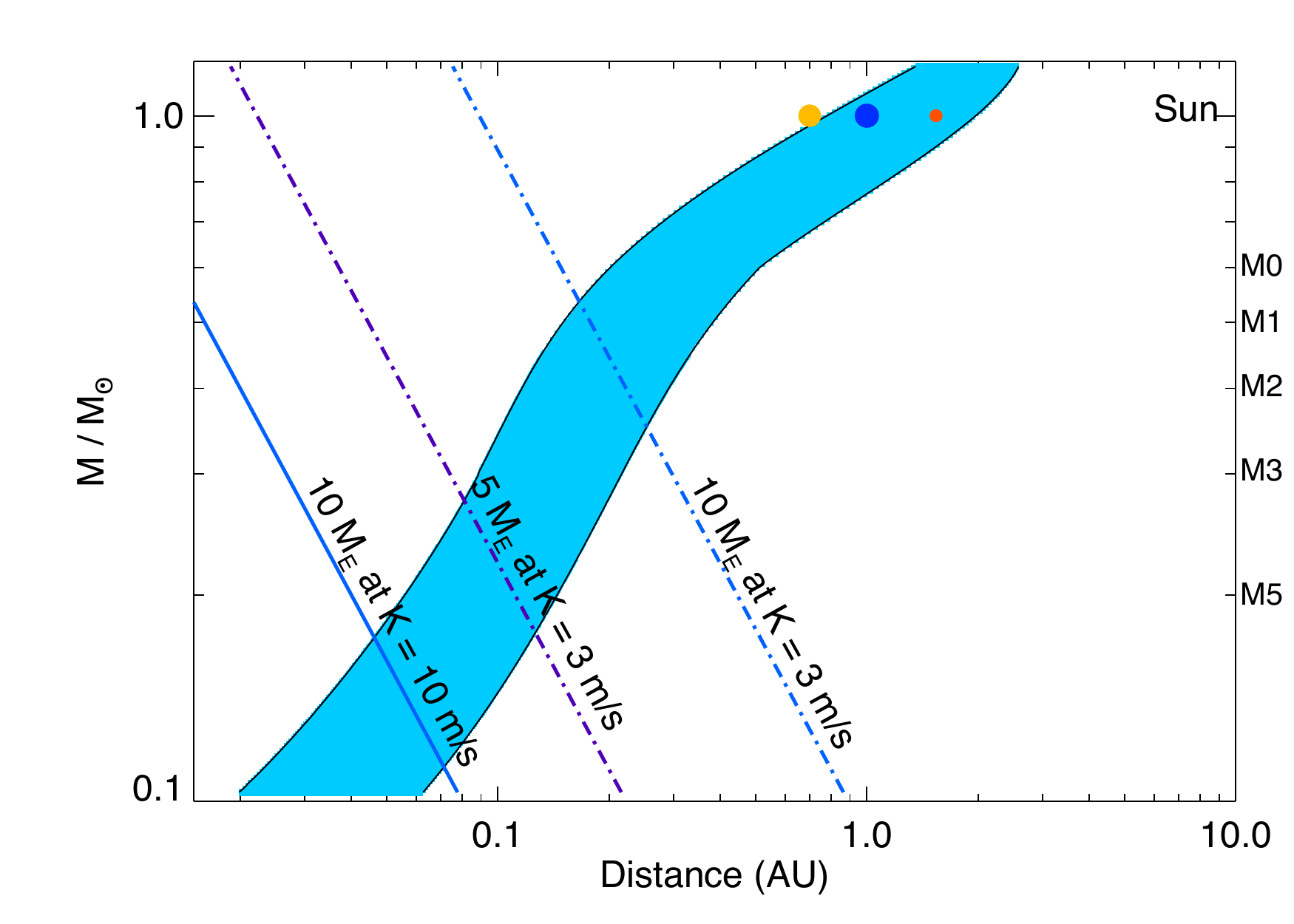}
\caption{The Habitable Zone around main sequence stars, and the velocity semi-amplitude of the Doppler wobble induced by 5 and 10 Earth-mass planets on the star. Venus, Earth and Mars are shown as colored dots.}
\label{fig:hz}
\end{center}
\end{figure}

\section{Targets Selection \& Survey Design}
There are only a limited number of mid-late M dwarfs bright enough for a precision RV survey, so we have begun to generate a target list and survey plan to enable HPF design choices to be made on a scientifically sound basis. The final target list will be comprised of $\sim 300$ stars from M4-M9 spectral types. Our simulations indicate that the HPF can achieve 3 m/s RV precision on a slowly rotating M4-M9 star with a S/N $>$ 150 per pixel in the extracted 1-d spectrum, which corresponds to a brightness limit of J=10 for a 15 minute exposure at the HET (assuming a 7m effective aperture and a 4\% total throughput). Figure 2 illustrates the search space opened up by the HPF, including the ability to
probe numerous M4/M5 and later stars, many of which may be beyond the reach of even a HPF clone on a 3-4m class telescope. The radial velocity information content available in stellar spectra degrades as the rotational velocity (vsini) rises \cite{2001A&A...374..733B}, so slower rotators are better
targets. To measure the rotational velocity of  a subset of potential targets we have previously acquired observations of M4-M7 stars with the HRS on the HET \cite{2009ApJ...704..975J}, and M6-M9 with NIRSPEC on Keck \cite{2012arXiv1207.2781D}.
When combining our work with measurements from the literature we find 200 M4-M9 stars with vsini
$<$12 km/s. These are shown in Figure 2 as blue circles. The figure also shows in grey dots the larger sample of star identified as M dwarfs from proper motion catalogs \cite{2011AJ....142..138L, 2005AJ....129.1483L}.

The HPF targets will be drawn from this large sample and will fulfill the following criteria:

\begin{itemize}
\item{Observable with the HET}
\item{J$<10$ (M4-­‐M8), J$<11$ (M8-­‐M9)}
\item{Vsini $<12$ km/s}
\end{itemize}

\begin{figure}  
\begin{center}
\includegraphics[width=6.5in]{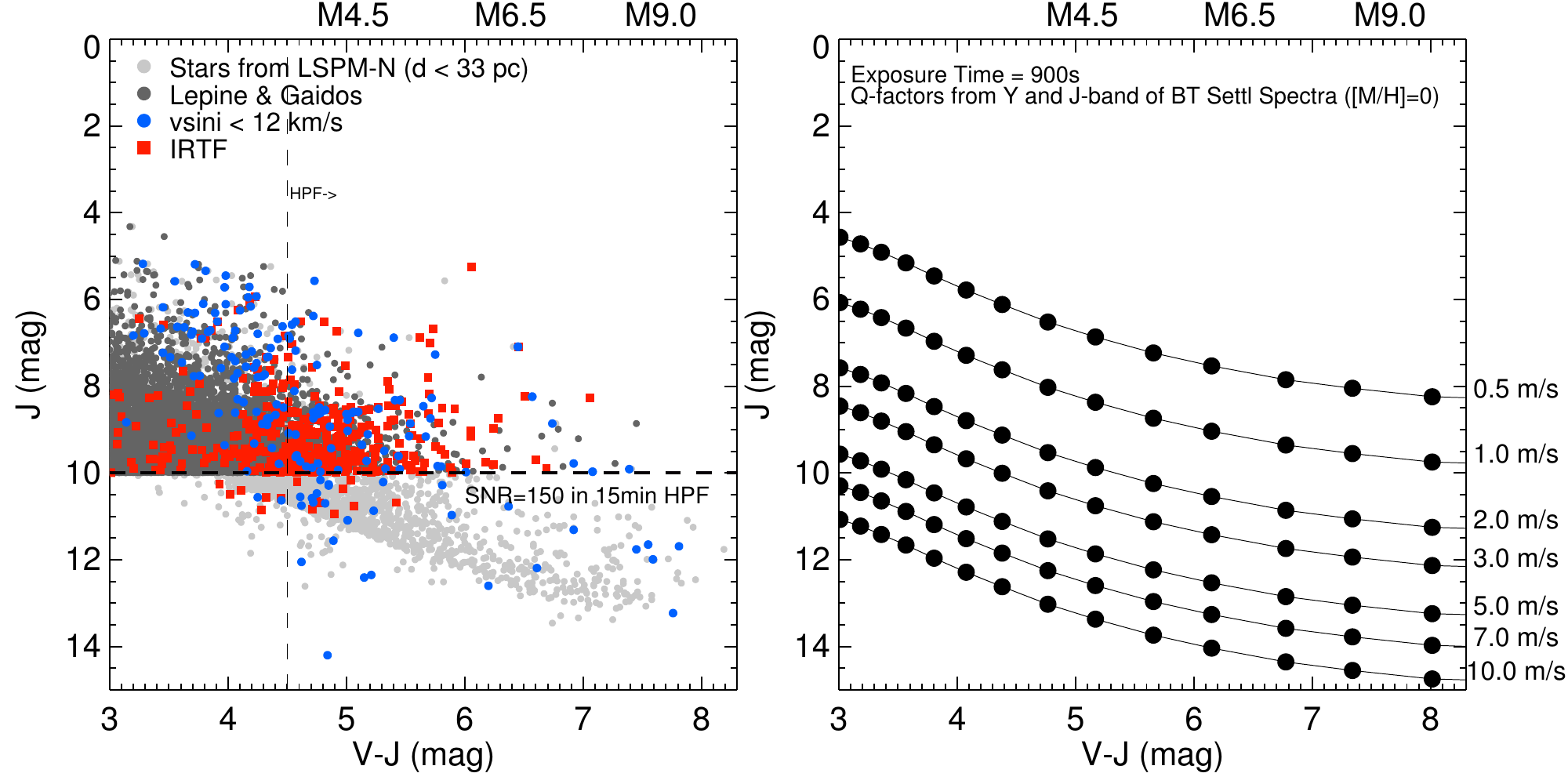}
\caption{Left: V-J color and J magnitudes of M stars. M stars from our
vsini $<$ 12 km/s sample are shown in blue circles, and those with IRTF Fe/H are red squares. The M dwarf catalogs are shown in grey. The horizontal dashed line corresponds to a S/N=150 per pixel expected with HPF on the HET. Right:  Simulations showing the expected theoretical limit of RV precision for an M dwarf observed for 15 minutes with the HPF in the Y \& J bands. Telluric absorption, detector noise, instrument effects etc. will reduce this theoretical precision.}
\label{fig:ffp}
\end{center}
\end{figure}

We caution that the calculated precisions shown in Figure 2 (right) are intrinsic photon-noise and information content limited precision for a slowly rotation target, calculated using  synthetic models (the BT-Settl models in this case). Real effects like read noise, tellurics, wavelength calibration, fast rotation etc. will degrade the RV precision. Synthetic models also do not adequately reproduces the lines seen in real M dwarf spectra, and the real information content in observed spectra could be quite different. The models do indicate that stellar metallicity (Fe/H) plays a key role. This and other aspects of our ongoing work on simulating the RV precision of HPF and better understanding the innate information content are presented in  Terrien et al. 2012 (these proceedings). Our ongoing NIR observations at the  NASA Infrared Telescope Facility (IRTF) have enabled us to develop a new empirical calibration technique for M dwarf metallicities \cite{2012ApJ...747L..38T} and ongoing surveys of $\sim~$2000 M dwarfs with the Sloan Digital Sky Survey (SDSS-III) APOGEE multi-object instrument \cite{2010SPIE.7735E..46W} will enable us to identify fast rotators and binaries. These ongoing efforts are designed to  yield a list of 500 suitable stars by the time of HPF commissioning, from which we will select 300 that span M4-M9 to optimize and fill the HET queue.

{\bf The HPF HET M dwarf Survey:} For an RV survey with HPF/HET we assume an average exposure
time of 15 minutes per star and a fixed overhead of 5 minutes per observation for the majority of targets, which corresponds to a S/N=150 per pixel at J=10 with our current assumptions of efficiency and effective telescope aperture.
For M7-M9 targets, which are typically much fainter, we will use exposures of 20-25 minutes. We use the history of published planets from optical RV surveys as a guide to estimate the number of observations needed to confidently identify a low mass planet. Figure 3, derived from the Exoplanet Orbit Database \cite{2011PASP..123..412W}, shows that with real-life effects like stellar jitter and systematics, optical RV surveys need $\sim$ 80--100  observations to reliably identify a planet with a semi-amplitude of 1-4 m/s. High cadence sampling is essential when the amplitude of the stellar reflex motion is comparable to the measurement precision RMS \cite{2008PASP..120..531C}. However, identifying giant planets, or unpromising sample members, requires many fewer observations. Low amplitude RV signals can be hard to detect due to stellar activity. Here the NIR (and choice of slow rotators) helps, since the RV signal induced by star spots is less than in the optical \cite{2010ApJ...710..432R}. The high resolution of the HPF also enables line-bisectors to be used to disentangle activity from planets. We will use the cascaded survey structure outlined in Table 1 to optimize our use of telescope time and to ensure the most promising sample members receive sufficient observations to detect the signal of a low-mass planet at 1-3 m/s. The initial 300
sample stars will be observed for five epochs each; 200 of the most promising will be observed with another 10 epochs. Two additional cuts will result in a total of 100 epochs of observation for the 50 most favorable stars. Observations will occur at short, intermediate, and long timescales to probe a variety of possible periods. This cascading technique mirrors the approaches used by successful optical surveys. The HPF survey will require over 2000 hours of HET time spread across 5 years, ie. more than 400 hours per year. The HET observes in queue-scheduled mode, which is ideal for temporal surveys since it enables optimal scheduling of observations. The large dark-time HET Dark Energy Experiment (HETDEX) will be operational at the HET in the early years of the HPF survey. The bright-time HPF M dwarf survey is an excellent complement to this.

\begin{figure}
\begin{center}
\includegraphics[width=4in, angle=90]{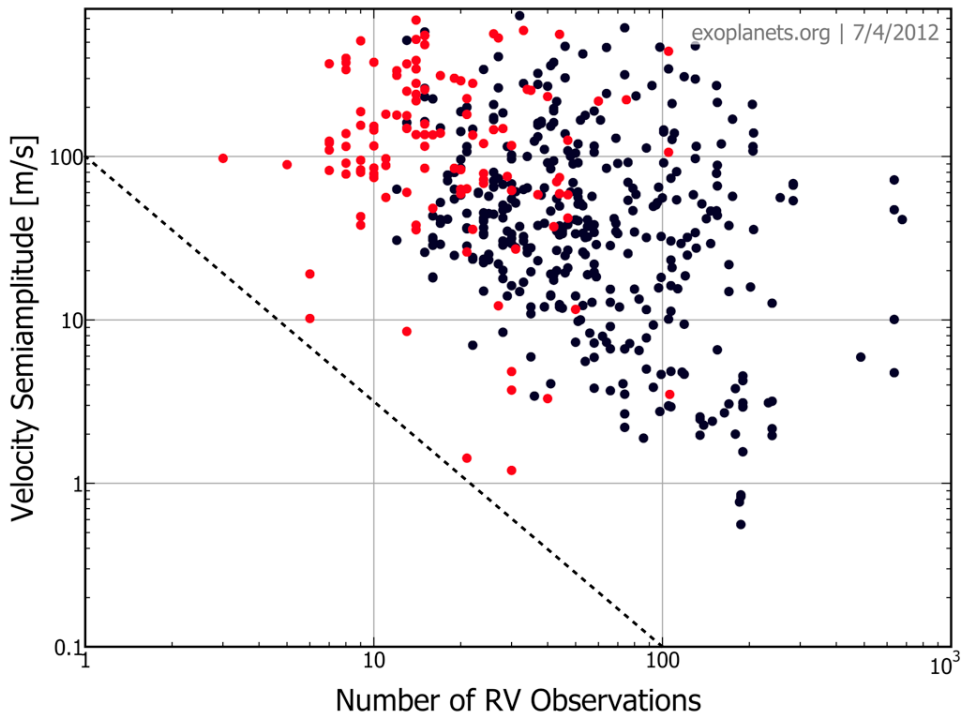}
\caption{Velocity semi-amplitude and number of published
RV observations for planets discovered with RV (black) and
transits (red).}
\label{fig:rv}
\end{center}
\end{figure}

\begin{deluxetable}{cc}
\tablewidth{0pt}
\tablecaption{HPF M Dwarf RV Survey Design }
\tablehead{
\colhead{Number of M stars}  & \colhead{Cumulate Number of RV Observations}
}
\startdata
300	&  5		\\
200  &  15		\\
100  &  30	 	\\
50    &  100		\\
\enddata
\label{tab:survey}
\end{deluxetable}

\section{HPF: Elements of Design Choice}
\label{sec:design}
A fiber-fed spectrograph can accept a simultaneous emission lamp calibration signal, enabling the
information rich  Y band (where no gas cells currently exist) to be used for RV
measurements. The HPF on the HET will be the only fiber-fed NIR spectrograph on a large 8-10 m
telescope capable of high-resolution, high-stability, and large simultaneous wavelength coverage.
Our instrument derives its heritage from the Gemini-commissioned PRVS instrument study, in which
Penn State was a major partner. PRVS won the Gemini instrument competition, but was cancelled by the
Gemini Board due to lack of funding. Our HPF design draws heavily on the PRVS design study
experience, as well as recent experience with the SDSS-III APOGEE spectrograph. Tests with our
prototype Pathfinder testbed in the lab and at the HET have demonstrated 10-20 m/s RV, mitigated risks,
and have clarified the path to 3 m/s, particularly in regard to calibration techniques. HPF on the HET
will enable fainter stars
to be observed, and ensuring that HPF will not be severely target limited even in an internationally competitive
milieu.  In this section we motivate key elements of the HPF design, current challenges, their solutions, and choices we need to make.

\subsection{Fiber Feed:}
 To ensure that most of the light is collected from the HET, even in seeing worse than the
0.9 arc second median EE50 at the HET site, a 300 $\mu m$ HPF fiber will subtend 1.7 arc seconds at f/3.65 with the new HET
wide field upgrade \cite{2010SPIE.7733E.129S}, while a 200 $\mu m$ HPF fiber will subtend 1.13 arc seconds. Focal ratio degradation will be minimal with this fast beam.
Achieving the design resolution of R=50,000 with a 200 mm beam and 3.3 pixel
sampling requires a 100-150 $\mu m$  slit. We will use either an image slicer or fiber slicer to collect the light and
reformat it into a long slit. Our quasi-Littrow spectrograph concept does introduce some variable slit tilt
across an order, but with an out-of-plane ($\gamma$) angle of only 0.5 deg, this slit effect is small. A hexagonal
array of seven 100 $\mu m$ fibers can collect $\sim$68\% of the light from an original 300 $\mu m$ fiber if the cladding
layer is kept thin. We have tested such fiber bundles in the Pathfinder development,
and find that for 1-2 m lengths throughput losses and cross-talk are minimal up to 1.7$\mu m$ even with a 1.1
clad-to- core diameter ratio. Fiber modal noise is also an important issue in the NIR, and we discuss our progress on this front in Section 4. A double scrambler that images the near field
output of one fiber to the far field output of another fiber can effectively minimize any shifts in the output
illumination profile  and mitigate the effect of imperfect guiding, focus drifts,
and HET pupil variations. The fiber link will also likely include a short stretch of octagonal fibers to enhance scrambling. The design of the HET leads to pupil illumination changes across a track, and scrambling is essential to maintain a stable PSF.  Along with mechanical agitation, we include a double scrambler as part of the
baseline concept. Vacuum feedthroughs for the fibers will employ the epoxy-based strategy successfully
demonstrated as part of the APOGEE  instrument build \cite{2010SPIE.7735E.207B}.

\subsection{Choice of Hawaii-2RG NIR Detector:}  
\begin{figure}
\begin{center}
\includegraphics[width=3.3in]{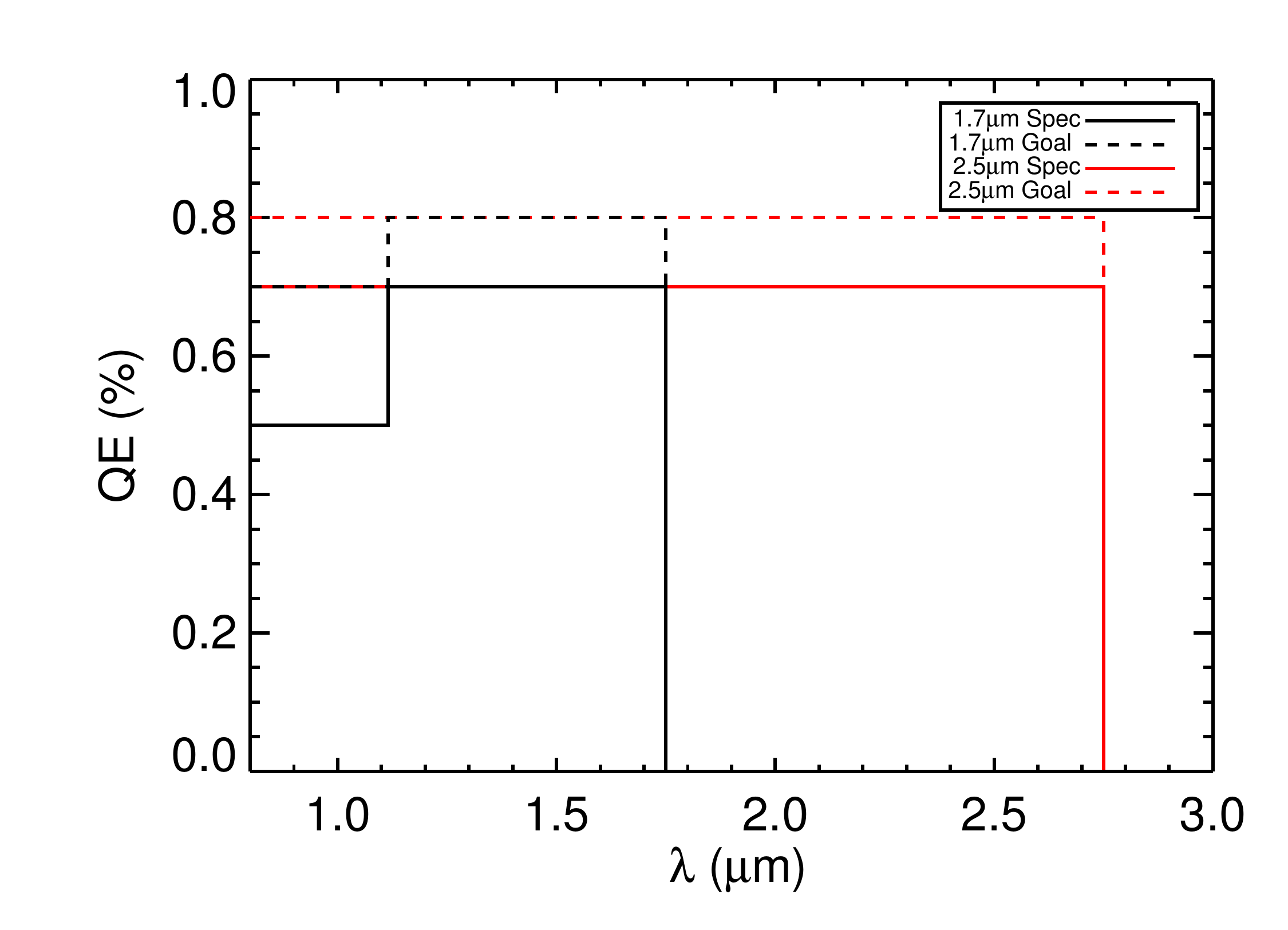} \includegraphics[width=3.3in]{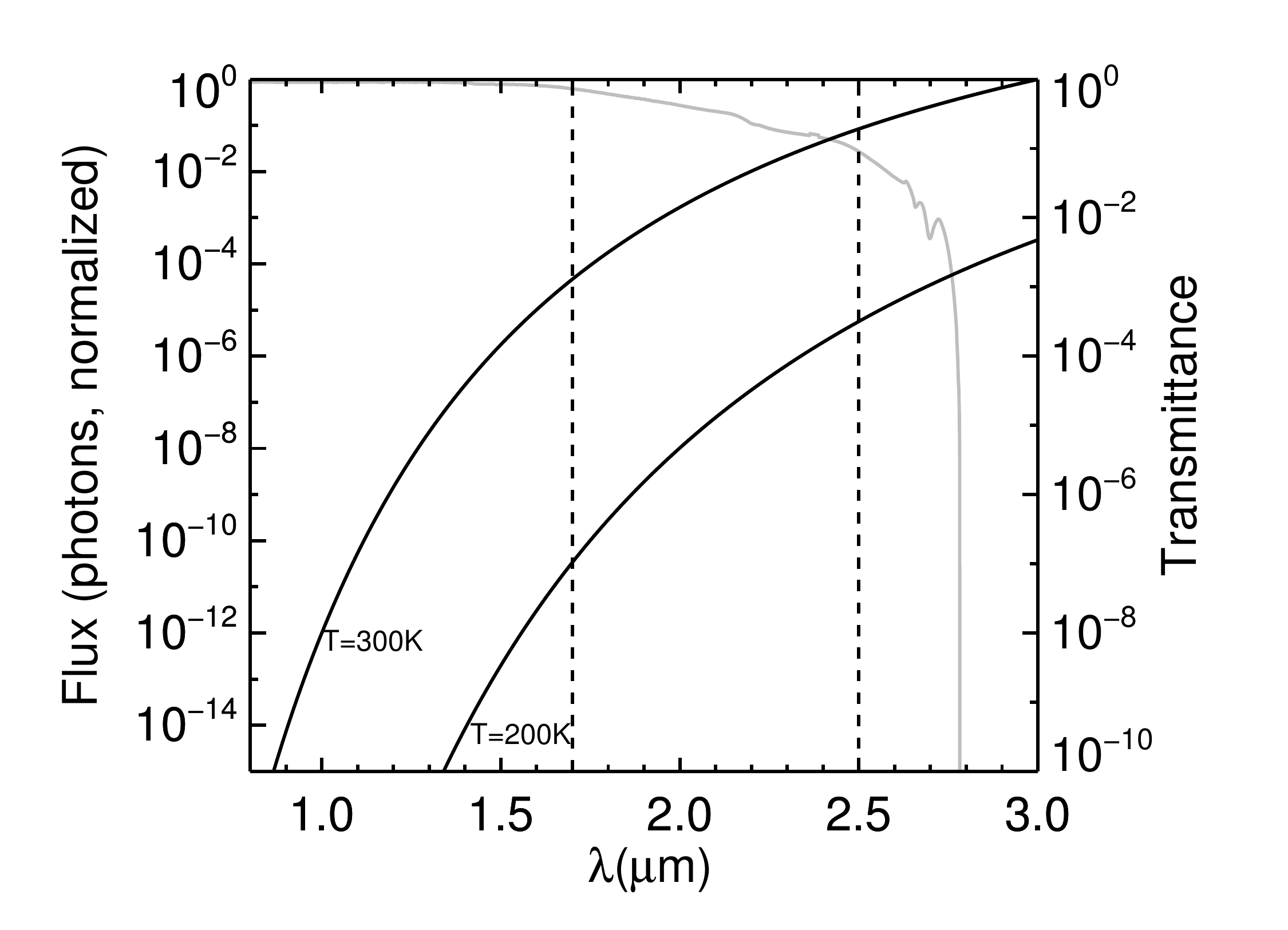}, 
\caption{(Left) QE, minimum spec. and goal, for the H2RG 1.75 (black lines) and 2.5 $\mu m$ (red lines) devices. (Right) Cumulative Flux from thermal background for 300K and 200K instrument temperatures. Dashed lines correspond to 1.7 and 2.5 $\mu m$. The grey line is the measured transmission through a 12mm block of PK50 glass.}
\label{fig:h2rg}
\end{center}
\end{figure}

The Teledyne Hawaii-2RG series of detectors is the design choice for HPF on account of their high quality, low persistence, low inter-pixel capacitance (IPC), and acceptable read noise. H2RGs are being used for precision photometry and spectroscopy, and our own tests with our Pathfinder spectrograph have shown that 10m/s RV precision can be achieved on the Sun\cite{2008PASP..120..887R}, and on starlight \cite{2012OExpr..20.6631Y}  with a Hawaii-1. The H2RGs are much improved successors to the Hawaii-1, and we expect that they will enable 1-3m/s precision with HPF. Our detailed simulations of the detector properties (Terrien et al. 2012, these proceedings) shows that dark current, IPC and normal persistence at the level expected in the modern arrays do not significantly degrade RV precision since a large number of pixels ($>10-20$) are used to sample each resolution element. Read noise at the level of 15-30 e/pixel does degrade the S/N and the final RV precision. This can be mitigated via a number of known read-out strategies like Fowler and up-the-ramp sampling at the expense of additional operational complexity. With HPF operating only in the Y \& J bands (and potentially in H in the future), there is no need for a fully cryogenic instrument operating at 80 K. With an instrument temperature below 200K the thermal background even in the H band is very low. A 1.75 $\mu m$ cut-off H2RG would be the ideal choice since no sensitivity in the K band is needed. This other option, a 2.5 $\mu m$ cut-off detector would require either cooling the instrument down to significantly lower temperatures to mitigate the thermal background (leading to additional cost and design complexity) or use of a thermal blocking filter to render the detector insensitive to wavelengths larger than 1.75 $\mu m$.  The 1.75 $\mu m$ would seem the obvious device (and is still the preferred choice for HPF), but the choice is complicated by  lower quantum efficiency (QE) and higher read noise of the 1.75 $\mu m$ devices. Figure \ref{fig:h2rg} illustrates the QE problem. The lattice structure of the HgCdTe material for the shorter wavelength cutoff devices is not as well matched to the CdZnTe substrate on which the detectors are grown, and the bandgap is higher, leading to  1.75 $\mu m$ devices that have poorer QE and read noise properties than their 2.5 $\mu m$ counterparts. For high  S/N spectroscopy, the read noise issue is potentially less important that the QE. As shown in Figure 4 the minimum Y band QE guaranteed by Teledyne for science grade H2RGs is only 50\% (goal of $>$ 70\%) for 1.75 $\mu m$, while it is 70\% (goal of $>$ 80\%) for the 2.5 $\mu m$ devices. Switching to using the  2.5 $\mu m$ devices is tempting given their superior qualities, but our experience with the Pathfinder spectrograph, and significant difficulties suppressing K band with the Hawaii-1 make us extremely wary of this approach. With H band tests designed to test a laser-comb we used a combination of a series of cryogenically cooled custom interference filters and PK50 glass inside the dewar to suppress light beyond $\sim1.7\mu m$\cite{2012OExpr..20.6631Y}. The thermal background was still many thousands of electrons/pixel for a 5 minute exposure. We attribute a significant fraction of this to pinholes even in high quality filters leading to light leaks. In addition filter performance is best on slow beams and HPF will have a fast focal ratio. The addition of the PK50 in our previous tests was very helpful, suggesting that while a 2.5$\mu m$ device formally cuts off beyond that wavelength, there is a long tail of low sensitivity to longer wavelengths, which leads to high background without cooling. Our experience cautions us against switching to a longer cutoff device without extensive tests and modeling. As shown in Figure 4, while cooling from 300K to 200K leads to a significant reduction in the thermal background, it still likely necessitates the use of a filter, or cooling further to $\sim 150$K. PK50 glass alone, while very effective beyond 2.5$\mu m$ is not useful to suppress 1.7-2.5$ \mu m$. We are not aware of any available glass does that, and very high quality custom interference filters will be needed, working at high input angles. We view this path as a very high risk at the moment, and the H2RG-1.7$\mu m$ is still the baseline device for HPF in spite of its shortcomings. Such a science grade array for HPF is being fabricated by Teledyne. 

\subsection{Spectrograph Optics:} 
\begin{figure}
\begin{center}
\includegraphics[width=5in]{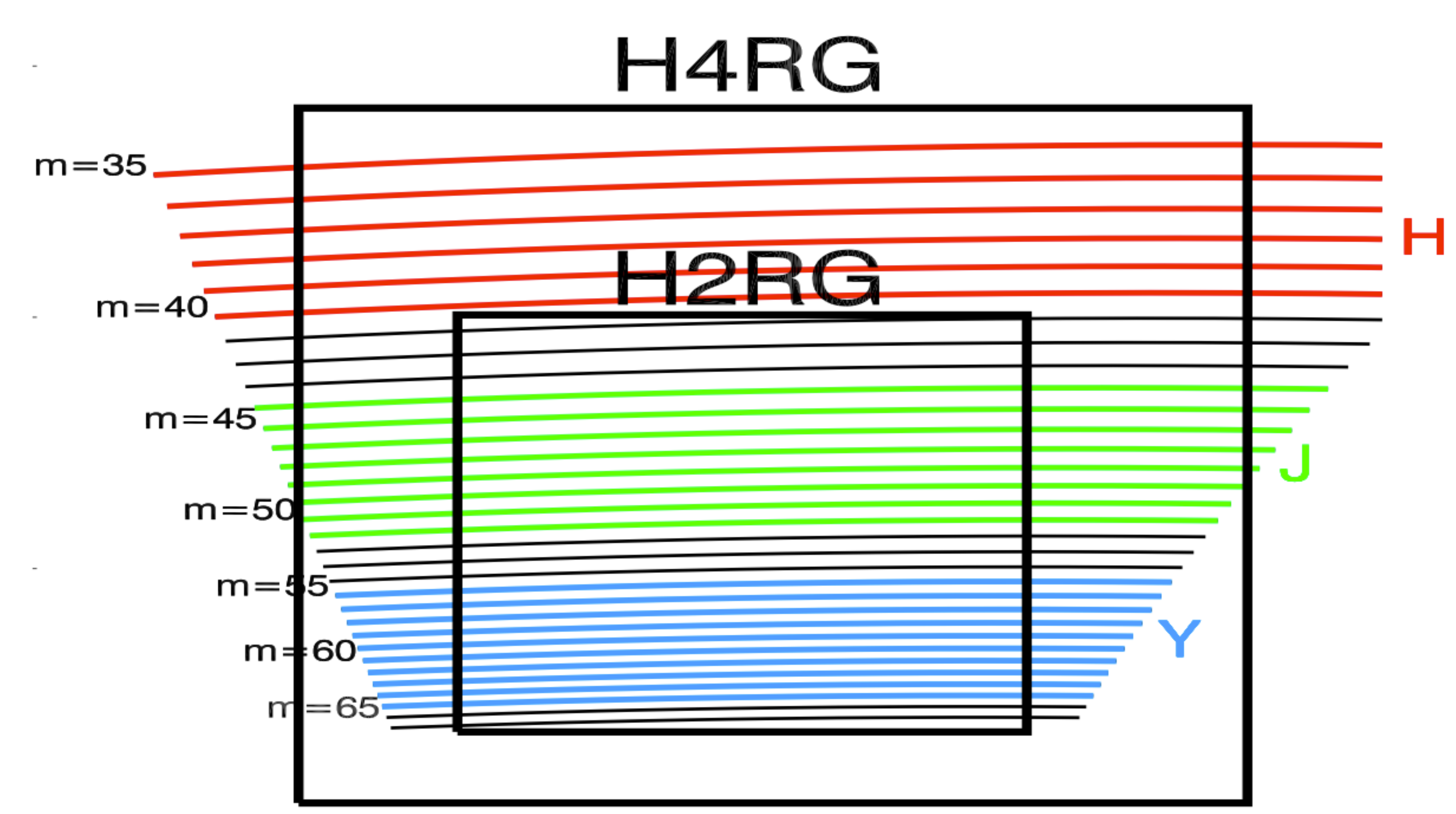} 
\caption{HZPF spectral coverage with H2RG array, and possible
future H4RG for once possible choice of cross-disperser. The echelle orders and Y, J , H bands are marked.}
\label{fig:coverage}
\end{center}
\end{figure}

We have explored various design options that meet the top-level science requirements. Silicon immersion gratings are ruled out because Si does not transmit well in the Y band where a lot of the M dwarf RV information content is. We have significant experience with dispersed fixed delay interferometry but conclude that this is a higher risk approach given our precision requirements. The baseline approach we adopt is the well proven quasi-Littrow white pupil design with a monolithic off-axis parabolic collimator, a 200 mm by 800 mm 31.6 g/mm replicated
mosaic grating from Newport Richardson Gratings Lab (RGL) blazed at 75 degrees, a fused silica grism cross disperser with a 110 g/mm 5.4 deg blaze epoxy grating replicated on it, and an f/2 refractive camera made with standard Schott and Ohara glasses (and all spherical surfaces). The white pupil enables good control of the scattered light and minimizes the size of the cross-disperser and camera apertures. The large mosaic echelle grating has been used in UVES and HARPS, and is a standard product from Newport RGL. The grism enables a very compact form factor for the spectrograph, and is conceptually similar to one of the same size manufactured by Newport RGL for HARPS. A low risk backup option (a 110 g/mm reflection grating from Newport) is available for the cross-disperser, though this increases the HZPF footprint. We are exploring both 150mm beam and a 200mm beam options, as well as the use of a volume phase holographic (VPH) grating cross disperser. This beam size beam enables all the refractive optics to be of reasonable size, while still enabling the resolution element to be sampled by 3-3.3 pixels on the H2RG yielding a resolution of R~50,000. Figure \ref{fig:coverage} shows the order format. HZPF will cover significant parts of Y (~75\%) \& J (~55\%) with the current H2RG, with possible extension to almost complete coverage of Y and J, and a significant part of H with a future H4RG (Figure 5). The choice of cross-disperser is still not fully defined, and we are still exploring choices of prisms and gratings. The issue is that prisms with standard glass materials do not typically provide sufficient dispersion in this wavelength regime, and gratings increase the instrument form factor.  The camera will be designed with a wide field for this possible
future upgrade. Figure 6 shows the optical design of the spectrograph. RMS spot sizes are all mostly 1 pixel for the H2RG. The monolithic collimator mirror, echelle grating, and the fold mirror will be Zerodur to reduce temperature sensitivity, and will all be gold coated to increase efficiency.

\begin{figure}
\begin{center}
\includegraphics[width=6in]{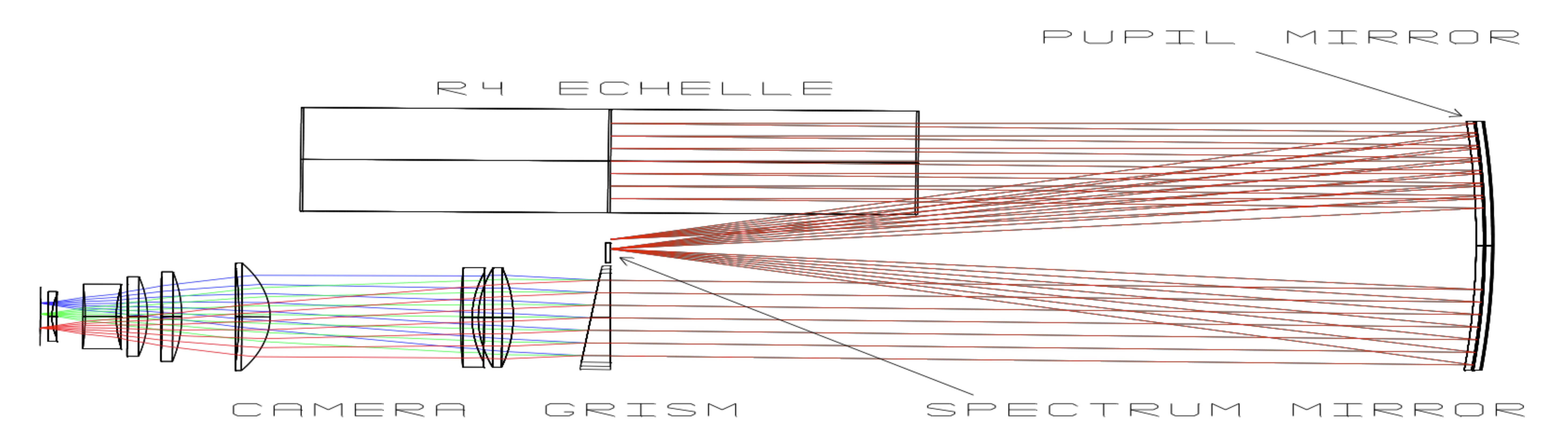} 
\caption{Optical Design Concept of the HZPF. Beam diameter is 150-200 mm, and RMS spot sizes are generally less that the pixel size (18$\mu$m) of the H2RG.}
\label{fig:optics}
\end{center}
\end{figure}

\subsection{Cryostat Design:} Spectrograph Cryostat: Successful implementation of our calibration technique requires high (few milli-Kelvin) thermal stability in the spectrograph. HARPS has demonstrated the approach; placing the entire instrument in a vacuum in a stable thermal environment. HZPF will be in a vacuum vessel, but also needs to be cooled so that ambient thermal radiation does not increase background. As part of the Gemini PRVS detailed study, we determined that the spectrograph components and enclosing structure need to be at ~200 K to mitigate the  thermal background. Our approach is based on recent successful development of the SDSS APOGEE instrument. The
cryostat is a stainless steel shell and cold structural components are constructed with 6061-T6 Aluminum
(Al). Basic parts consist of a monolithic lightweight Al 6061-T6 coldplate to hold the optical components. To minimize drifts due to weight changes, the  liquid nitrogen (LN2) will be mechanically isolated from the cold plate itself. Active thermal shielding surrounds
the assembly. The thermal shielding, designed to minimize radiative loads, is a 3003 Al shell wrapped in
multi-layer insulation (MLI). The LN2 tank, which will act as the heat sink, is a tube with welded ends,
and can hold over 90 liters of LN2, yielding a hold time for HPF of over a week. Radiation and custom-sized
thermal straps couple the LN2 heat sink to the cold plate and the spectrograph. These, combined
with precision-controlled active heating, ensure that spectrograph temperature stability is maintained at
the design temperature of 200K, with stability of $<$10 mK long term and a few mK overnight. This temperature regime is cold enough that thermal background is minimized with a 1.7$\mu m$ H2RG, while not being fully cryogenic (which increases cost and risk by requiring special glasses and a monolithic ruled grism/gratings). The H2RG detector itself will be cooled with LN2 and thermally controlled to mK levels as well. The use of a low power SIDECAR ASIC controller will minimize the temperature fluctuations. Existing and planned instruments already require a significant quantity of LN2 be maintained at the HET, which is already planned as part of HETDEX. The top of the vacuum chamber will be removable for access to the optical support structure. Figure 7 shows the concept design of the cryostat. 

\begin{figure}
\begin{center}
\includegraphics[width=6in]{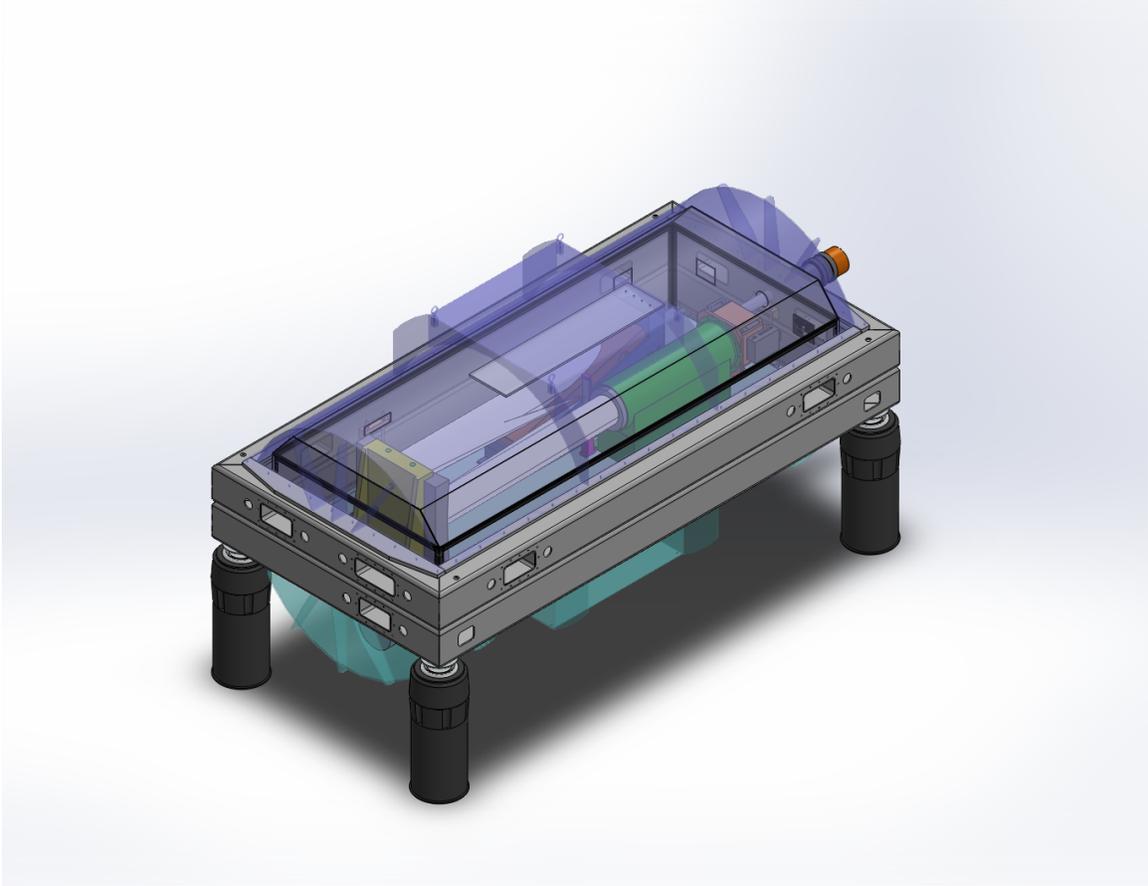} 
\caption{Concept design for the HPF Cryostat}
\label{fig:optics}
\end{center}
\end{figure}

\subsection{Wavelength Calibration Source:} 
Thorium-argon lamps (Th/Ar) that are widely used as a wavelength calibration source in the optical are not as useful in the NIR since thorium has very few strong lines in this region. The Argon lines, while extremely bright, are susceptible to pressure shifts, and are not stable enough for HPF's calibration needs.   We have demonstrated the utility of uranium neon (U/Ne) hollow-cathode lamps in the NIR, and have published derived line-lists using a Fourier Transform Spectrograph (FTS) \cite{2011ApJS..195...24R}, as well as a spectroscopic atlas calibrated with a laser frequency comb \cite{2012ApJS..199....2R}. U/Ne hollow-cathode lamps are commercially  available from Photron, and are the baseline calibration source for HPF to achieve its requirement of 3m/s RV precision.

The U/Ne lamp, while superior to Th/Ar for the NIR, still has unattractive features like lines blending, inhomogenous line coverage and intensity etc. at the HPF's spectral resolution. A NIR laser frequency comb (LFC) would significantly enhance the HPF radial velocity precision and accuracy.  Our ongoing collaboration between Penn State, NIST Boulder, and the University of Colorado  has already demonstrated the first on-sky radial velocities with a NIR laser comb using the Pathfinder testbed \cite{2012OExpr..20.6631Y, 2012ApJS..199....2R}. These tests were conducted in the H band, but at a spectra resolution very similar to that of HPF. Our collaboration is now moving towards designing and testing a Y \& J band LFC for use with HPF. While this effort is still unfunded, we are optimistic that an LFC can be designed and built for HPF by the time of the survey start. Alternate development paths to create stable Fiber Fabry Perot calibration devices to use with HPF during alignment and testing are presented in Halverson et al. 2012 (these proceedings), and described briefly in Section \ref{sec:RD}.
 
\section{Ongoing R\&D Projects to Mitigate Risk}

While experiments with our Pathfinder testbed have mitigated risk on many aspects of the HPF design, there are obviously many key differences between an exploratory warm-bench spectrograph achieving 10 m/s RV precision, and a cooled facility instrument whose goal is 1m/s.  Therefore, we have initiated a number of R \& D projects to mitigate risk in various aspects of the HPF design.

\label{sec:RD}
\subsection{Modal Noise Mitigation: } Modal noise, caused by the finite number of  electromagnetic modes propagating in an optical fibers, can introduce noise effects in high S/N spectra \cite{2001PASP..113..851B},  causing systematic errors in measured RVs. This is a more serious problem for NIR spectrographs than for optical spectrographs since the longer wavelengths populate fewer modes in the fiber, leading to more modal noise. The situation is exacerbated when coupling the single-mode fiber output of the LFC to the highly multi-mode fiber input of a spectrograph. Our experiments with Pathfinder and an LFC at the HET were affected by this, even though we used an integrating sphere and a fiber agitator.  A similar test at Penn State using a 1550nm laser as a proxy for the comb has shown that bulk motion of the fiber is needed to effectively fix modes and mitigate modal noise. Fig 8 shows the power spectrum of an image formed by illumination of a 200$\mu m$ core static fiber,  using an integrating sphere with a high frequency agitator, and using bulk hand motion.  Mechanical agitators are now being designed to replicate the effects of this bulk hand motion. {\it We find that including a 2m stretch of octagonal fiber does {\bf not} improve the modal noise distribution}. Our ongoing tests on modal noise and its mitigation are presented in greater detail by McCoy et al. 2012 (these proceedings), but it is worth to reiterating here that we have succeeded in suppressing the  high frequency power in speckles to 100x better than the Pathfinder+LFC tests at the HET. Ongoing tests with double scramblers, octagonal fibers, and mechanical agitators will enable us to develop a robust input feed HPF for that performs  a high degree of scrambling and  modal mixing.

\begin{figure}
\begin{center}
\includegraphics[width=5.5in]{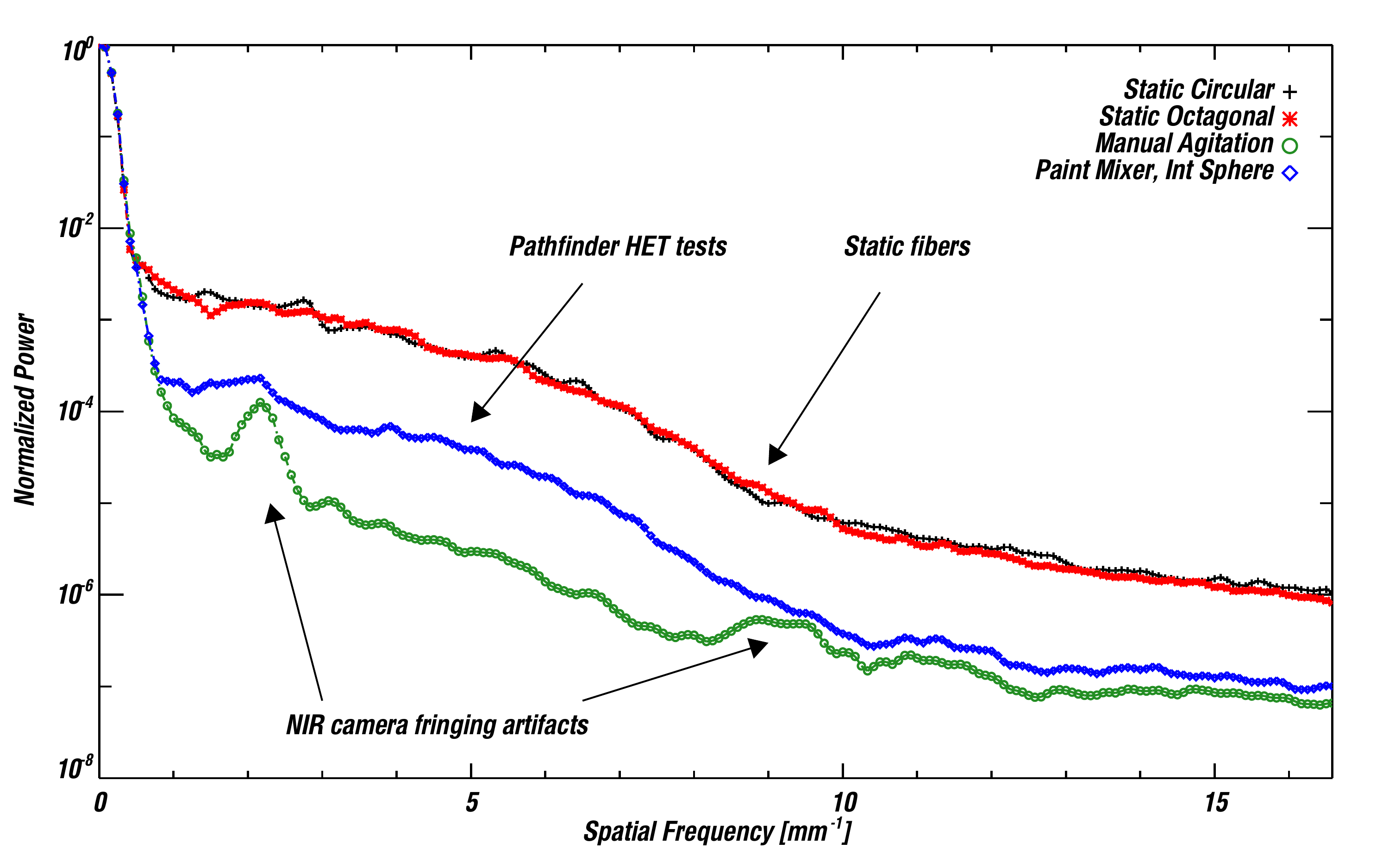} 
\caption{Power spectrum of modal noise through static and agitated fibers. Bulk agitation by hand significantly outperforms the use of just an integrating sphere. }
\label{fig:modal}
\end{center}
\end{figure}

\subsection{New Calibration Source: A Fiber Fabry Perot}
Testing Pathfinder with the NIST H band laser comb has  demonstrated the utility of having available a source that yields a dense grid of bright sharp stable lines. Such a source would be very useful for alignment, focusing, modal noise testing, and to track instrument stability. Clearly the LFC is all these things, but is an expensive resource for general lab use. As a less expensive alternative we have developed and tested a new Fiber-Fabry Perot (FFP) calibrator that resembles (to the spectrograph) an LFC,  can be stable to $\ll$ 3m/s, but does not have intrinsic frequency of each line known, except by comparison with an U/Ne or laser. Fig 9 shows high resolution scans of the FFP with the NIST Fourier transform spectrograph. Since it is constructed with a single mode fiber (Corning SMF-28) a supercontinuum source is needed to pump enough light through. Results of detailed tests of this H band FFP are presented in Halverson et al. 2012 (these proceedings), and we are planning to develop an FFP that will work in the Y\& J bands for use with HPF.

\begin{figure}
\begin{center}
\includegraphics[width=5.5in]{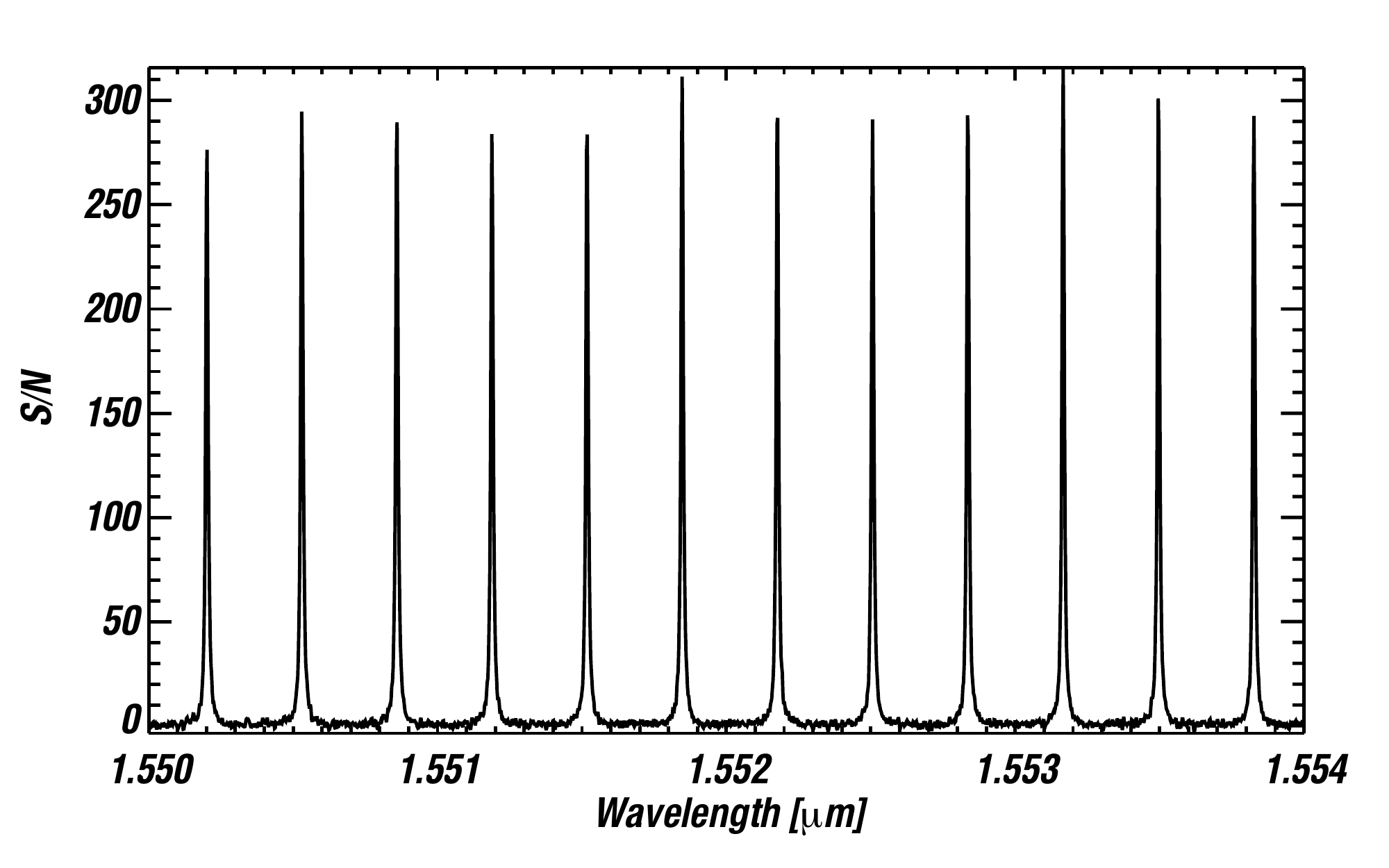} 
\caption{FTS scans of a small wavelength region from the FFP output. See paper by Halverson et al. in these proceedings for details on the FFP, and tests with it}.
\label{fig:modal}
\end{center}
\end{figure}

\subsection{Cooldown Tests:}
Maintaining HPF at ~170-200K, and at a temperature of a few milli Kelvin around that set-point likely requires an active control mechanism.  Colleagues at the University of Virginia (UVa) will conduct tests on their existing test cryostat to simulate these conditions, and attempt to achieve a high level of thermal control. Parts are being machined and the first set of cool down tests will be conducted this summer. The results of these tests will help guide us to a better conceptual design, and reveal risks inherent in this technique. To date the most stable instruments (e.g.. HARPS ) have used active control of the room, and let the vacuum chamber and insulation dampen changes to have very stable temperatures inside the chamber. However this is much harder for IR instruments that may have to stabilize the temperature actively if very precise control is need. Such active stabilization is currently the baseline for HPF.

\subsection{Echelle Grating Epoxy \& Cooldown}
The grating baselined for HPF is a 200 x 800 mm R4 mosaic from Newport RGL. All current monolithic mosaics are epoxy replicated on either Zerodur or Aluminum, and we have chosen the Zerodur substrate.  These are generally used at room temperature though, while Aluminum masters are used for full cryogenic applications.  HPF, working at $\sim200$K could benefit from the stability of the Zerodur, and the epoxy is likely to not be adversely affected by this level of cooldown. This, however, is a  risk that we are trying to address by 

\begin{itemize}
\item{Working Closely with Newport RGL on determining if previous customers have used such systems at low temperatures.}
\item{Acquiring a small test grating and run it through multiple cooldown cycles using the UVa test dewar we have described above.}
\end{itemize}

\section{Conclusion}
Our goal here is to present the motivation, survey  targets, conceptual design, and key properties of the Habitable-zone Planet Finder. A number of ongoing R\&D development efforts have also been highlighted, that demonstrate control of modal noise, development of calibration sources, and are described in more detail in companion papers in these proceedings. While building on a decade of impressive development in precision RV in the optical, HPF will face additional (and sometimes unexpected) challenges due to its need to operate in the NIR. We intent to publish our findings and instrument design and performance results in future SPIE proceedings.

\acknowledgments{This work was partially supported by the Center for Exoplanets and Habitable Worlds, which is supported by the Pennsylvania State University, the Eberly College of Science, and the Pennsylvania Space Grant Consortium. We acknowledge support from NSF grant AST-1006676,  AST-1126413, the NASA Astrobiology Institute (NAI), and PSARC. This research was performed while SLR held a National Research Council Research Associateship Award at NIST.}

\bibliography{HPF_SPIE}   %>>>> bibliography data in report.bib
\bibliographystyle{spiebib}   %>>>> makes bibtex use spiebib.bst

\end{document}